\documentclass[twocolumn]{aastex6}
\usepackage[utf8]{inputenc}
\usepackage{amsmath}
\usepackage{booktabs}
\newcommand{\red}[1]{{#1}}
\newcommand{\blu}[1]{{#1}}
\newcommand{\vsini}{\ensuremath{v \sin{i}}}

\newcommand{\kms}{km~s$^{-1}$}

\begin{document}

\title{Rotationally Driven Ultraviolet Emission of Red Giant Stars}
\author{Don Dixon\altaffilmark{1,2}
Jamie Tayar\altaffilmark{3,4},
Keivan G. Stassun\altaffilmark{2,1}}
\altaffiltext{1}{Department of Physics, Fisk University, Nashville, TN 37208, USA}
\altaffiltext{2}{Department of Physics and Astronomy, Vanderbilt University, Nashville, TN 37235, USA}
\altaffiltext{3}{Institute for Astronomy, University of Hawaii, Honolulu, HI 96822, USA}
\altaffiltext{4}{Hubble Fellow}

\begin{abstract}
\blu{Main sequence stars exhibit a clear rotation-activity relationship, in which rapidly rotating stars drive strong chromospheric/coronal ultraviolet and X-ray emission. While the vast majority of red giant stars are inactive, a {\bf few percent} exhibit strong ultraviolet emission.}
\blu{Here} we use a sample of 133 red giant stars \blu{observed by SDSS APOGEE and {\it GALEX\/}} to \blu{demonstrate an} empirical relationship between $NUV$ excess and rotational velocity ($v\sin i$).
{Beyond this simple relationship, we find that $NUV$ excess also correlates with rotation period and with Rossby number in a manner that shares broadly similar trends to those found in M~dwarfs, including activity saturation among rapid rotators. Our data also suggest that the most extremely rapidly rotating giants may exhibit so-called ``super-saturation", which could be caused by centrifugal stripping of these stars rotating at a high fraction of breakup speed.} 
\blu{As an example application of our empirical rotation-activity relation, we demonstrate that the $NUV$ emission observed from a recently reported system comprising a red giant with a black hole companion is fully consistent with arising from the {rapidly rotating red giant in that system}.} 
Most fundamentally, our findings suggest a common origin of chromospheric activity in rotation {\bf and convection} for cool stars from main sequence to red giant stages of evolution.

\end{abstract}

\maketitle

\section{Introduction}

There is a clear correlation between rotation and activity in stars \citep[e.g.][]{Kraft1967,Noyes1984}. Much work has been done to understand the reason for this relationship, including its relationship to stellar magnetism and the evolution of the stellar dynamo \citep{Wright2011}. 
Understanding these connections between stellar activity and rotation have become particularly important in the context of stellar and planetary evolution. The change in activity in low-mass stars has been used as a diagnostic of age \citep{MamajekHillenbrand2008} that has been suggested to be useful over a wider range of time than rotation alone \citep{MetcalfeEgeland2019}. In addition, the irradiation of young planets by high energy photons has been suggested to substantially alter the atmosphere \citep[see, e.g.,][and references therein]{Gaudi2017} and even change the location of the habitable zone \citep[e.g.,][]{Fossati2018}. 

Indeed, the connection between rotation and activity has a wide range of astrophysical applications. Work has been done to explore this relationship in solar-type stars \citep[e.g][]{Findeisen2011}, and M-dwarf stars \citep{2016MNRAS.463.1844S}. Additionally, it is clear that it evolves with stellar age, because stars spin down as they age \citep{MamajekHillenbrand2008}; it can therefore be used as a stellar age chronometer, and is not particularly affected by the presence of planets once selection effects are taken into account \citep{France2018}. 

However, these explorations have focused on dwarf stars; our understanding of rotational evolution in giants has therefore implicitly assumed similarity to dwarfs. Models of mass and angular momentum loss in giants through magnetized winds \citep{CranmerSaar2011}, for example, rely on assumptions about the relationship between X-ray flux, photospheric filling factor of open magnetic flux tubes, and rotation rate, all of which are calibrated primarily on dwarf stars. Similarly, the use of spots and activity to infer the rotation periods of giants has assumed that they have similar stability and lifetimes to spots on dwarf stars \citep{Ceillier2017}. These assumptions require testing, which has been challenging due to the fact that red giants are overwhelmingly slow rotators; only a small fraction of red giants are known to be active \citep[e.g.,][]{Ceillier2017}. 

Therefore, in this work, we identify a sample of stars with which the relationship between rotation and $NUV$ activity can be tested. We demonstrate that there is indeed a correlation between $NUV$ excess and rotation for giant stars, discuss how this correlation relates to relationships for dwarf stars, and discuss how it can be used to further our understanding of the structure and evolution of giant stars.

The structure of the paper is as follows. In Section~\ref{data_methods} we describe our sample selection for the study and our method for measuring ultraviolet excess. In Section~\ref{results} we present the correlation results between ultraviolet excess and rotation. Section~\ref{discussion} discusses potential implications and applications of our results. Finally, Section~\ref{conclusion} summarizes our key conclusions.

%
%
%
%
%











\section{Data and Methods}
\label{data_methods}

In this section, we describe the data that we use and our methods of analysis. The final observed and derived data that form our results in Section~\ref{results} are summarized in Table~\ref{tab:data}. 

\begin{table*}
    \centering
        \begin{tabular}{llccccccccccc}
        \toprule
        {} &           APOGEE ID &  LOCATION ID &        $T_{\rm eff}$ &  $\log g$ &  $v\sin i$ &  $\sigma_{v\sin i}$ &      $J$ &   $\sigma_J$ &     $K_S$ &  $\sigma_{K_S}$ &    $NUV$ &  $\sigma_{NUV}$ \\        \midrule
        0 &  2M08111162+3255049 &         4103 &  4772.47 &  2.45 &   4.27 &       1.56 &   8.78 &  0.02 &   8.14 &  0.02 &  16.17 &   0.02 \\
        1 &  2M10513105-0103246 &         4236 &  4624.71 &  3.18 &  11.85 &       0.50 &   9.71 &  0.03 &   9.02 &  0.03 &  19.09 &   0.02 \\
        2 &  2M07360651+2114107 &         4147 &  4798.02 &  3.05 &   9.51 &       0.38 &   9.30 &  0.02 &   8.67 &  0.02 &  18.01 &   0.03 \\
        3 &  2M00013362+5549387 &         4264 &  4768.57 &  3.09 &   8.80 &       0.76 &   8.94 &  0.03 &   8.08 &  0.03 &  18.53 &   0.09 \\
        4 &  2M08363324+1516597 &         4534 &  4537.72 &  3.06 &  12.56 &       0.09 &   8.09 &  0.03 &   7.38 &  0.03 &  17.01 &   0.03 \\
        5 &  2M18014184+6011596 &         4526 &  4843.05 &  3.23 &   3.66 &       2.71 &  11.99 &  0.02 &  11.43 &  0.02 &  19.88 &   0.10 \\
        6 &  2M12285832+1351453 &         4218 &  4622.74 &  3.15 &  11.04 &       0.86 &   9.85 &  0.02 &   9.20 &  0.02 &  18.80 &   0.02 \\
        7 &  2M11024366-0544050 &         4239 &  4974.92 &  3.24 &   4.44 &       3.14 &  12.14 &  0.03 &  11.57 &  0.03 &  20.20 &   0.08 \\
        8 &  2M12285645+1508139 &         4218 &  4419.97 &  3.29 &  11.49 &       0.00 &  11.88 &  0.02 &  11.17 &  0.02 &  20.11 &   0.04 \\
        9 &  2M08033852+7742081 &         4529 &  4845.88 &  3.12 &   1.98 &       1.58 &  12.32 &  0.02 &  11.74 &  0.02 &  21.27 &   0.31 \\
        \bottomrule
        \end{tabular}
    \caption{{Stellar parameters used to derive $NUV$ excess. The full table is provided in machine-readable form; a portion is shown here for guidance regarding its format and contents.}}\label{tab:data}
\end{table*}

\subsection{APOGEE Spectra}
\label{apogee_spectra}
\begin{figure}[!ht]
\centering
\includegraphics[width=\linewidth]{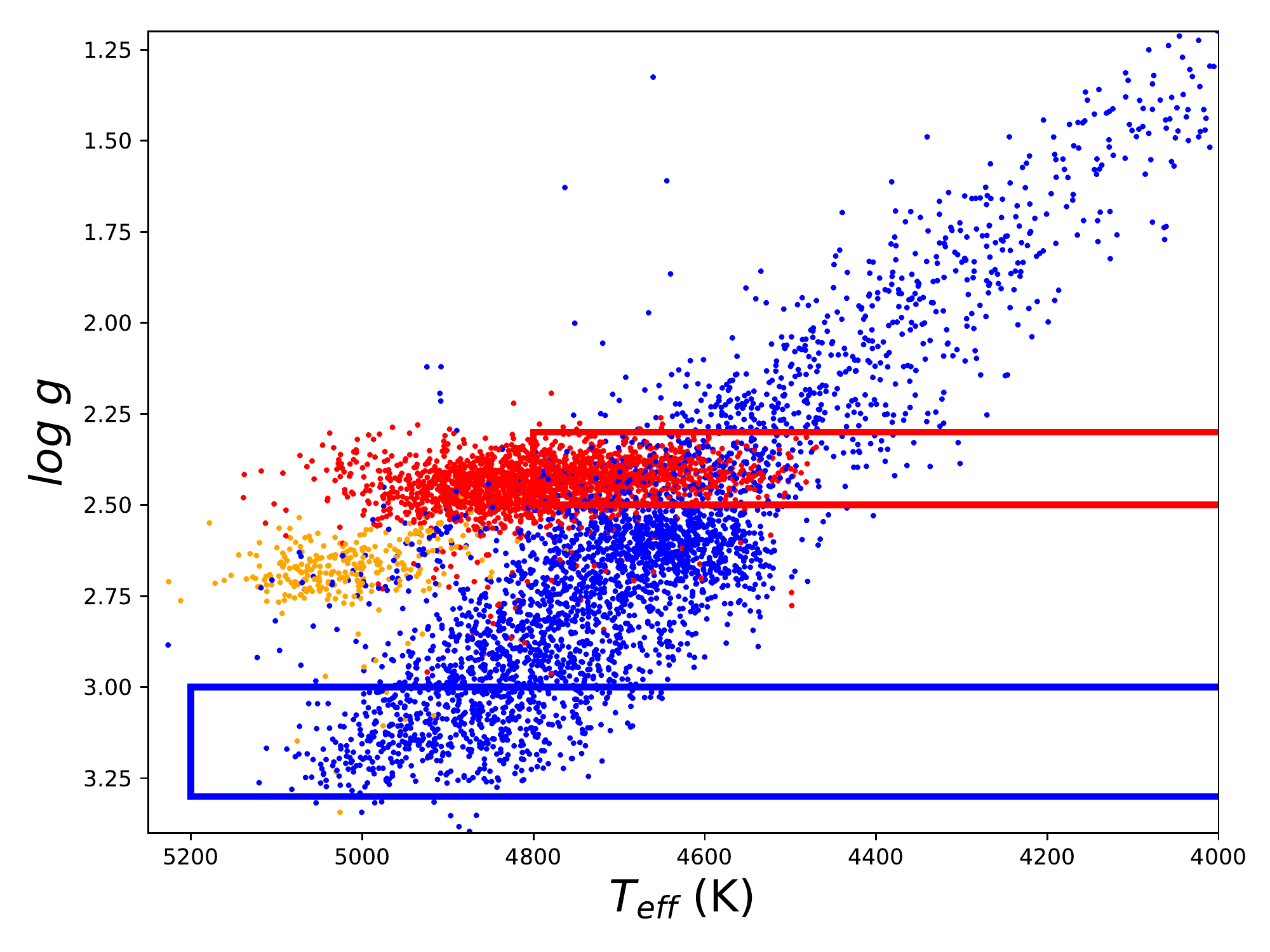}
\caption{
Observational HR histogram of APOKASC catalog red giant stars (blue symbols), red clump stars (red), and secondary clump stars (orange). Rectangular blue and red lines indicate our choice of cutoffs for RGB and RC evolutionary states respectively. This demonstrates our choice of cutoffs \red{effectively captures RC stars across the width of the entire giant branch, while avoiding contamination from secondary clump stars.} 
}
\label{fig:apokasc_params}
\end{figure}

To identify the red giant stars used in this analysis and to measure their rotation velocities, we rely on spectroscopic observations from Data Release~14 \citep{dr14} of the APOGEE survey \citep{Majewski2017}. APOGEE is a part of the Sloan Digital Sky Survey~IV \citep{Blanton2017} taking $H$-band spectra on the 2.5~m Sloan Digital Sky Survey telescope \citep{Gunn2006} for stars across the galaxy. These spectra are analyzed using the ASPCAP pipeline \citep{Nidever2015, GarciaPerez2015}, and the resulting gravities, temperatures, and abundances are calibrated using open clusters and field stars as discussed in \citet{Meszaros2013} and \citet{Holtzman2018}. Because we are particularly interested here in checking whether stellar evolutionary state affects the UV excess, we use the catalog of stars with asteroseismic evolutionary states published in \citet{Pinsonneault2018} to identify a locus of lower red giant branch (RGB) stars, and a locus of red clump (RC) stars. Specifically, we adopt $T_{\rm eff}< 5200$~K and $3.0 < \log g < 3.3$, and $T_{\rm eff}< 4800$~K and $2.3 < \log g < 2.5$, as filters for the RGB and RC evolutionary states, respectively. We compare these choices to the well characterized APOKASC catalog and find they accurately encompass the distributions of the targeted stellar types (Figure~\ref{fig:apokasc_params}).

\blu{Applying our choices as parameter cutoffs, we check for RGB and RC stars from APOGEE in the Tycho-Gaia Astrometric Solution (TGAS) catalog and a separate dataset of radial-velocity (RV) variable APOGEE stars taken from \citet{Badenes2018} and use them to form two separate samples of giants. The evolutionary makeup of giants in our TGAS and RV samples are markedly different: whereas the TGAS sample is predominantly RC stars \red{($\sim$59\%)}, the RV sample is predominantly RGB stars \red{($\sim$86\%)}. This is consistent with the expectation that the rate of binarity, as indicated by the rate of RV variables, decreases as stars evolve from the RGB to RC phase \citep{2020arXiv200200014P}. 
In the analysis that follows, we treat the TGAS sample stars as putative singles, whereas we treat the RV variable sample as putative binaries (where the putative binary companion to the observed red giant is not directly seen; i.e., the red giant is a single-lined binary); hereafter we refer to these samples as ``Field" and ``RVvar/Binary", respectively.}

\begin{figure*}[!ht]
\centering
\includegraphics[width=.85\paperwidth]{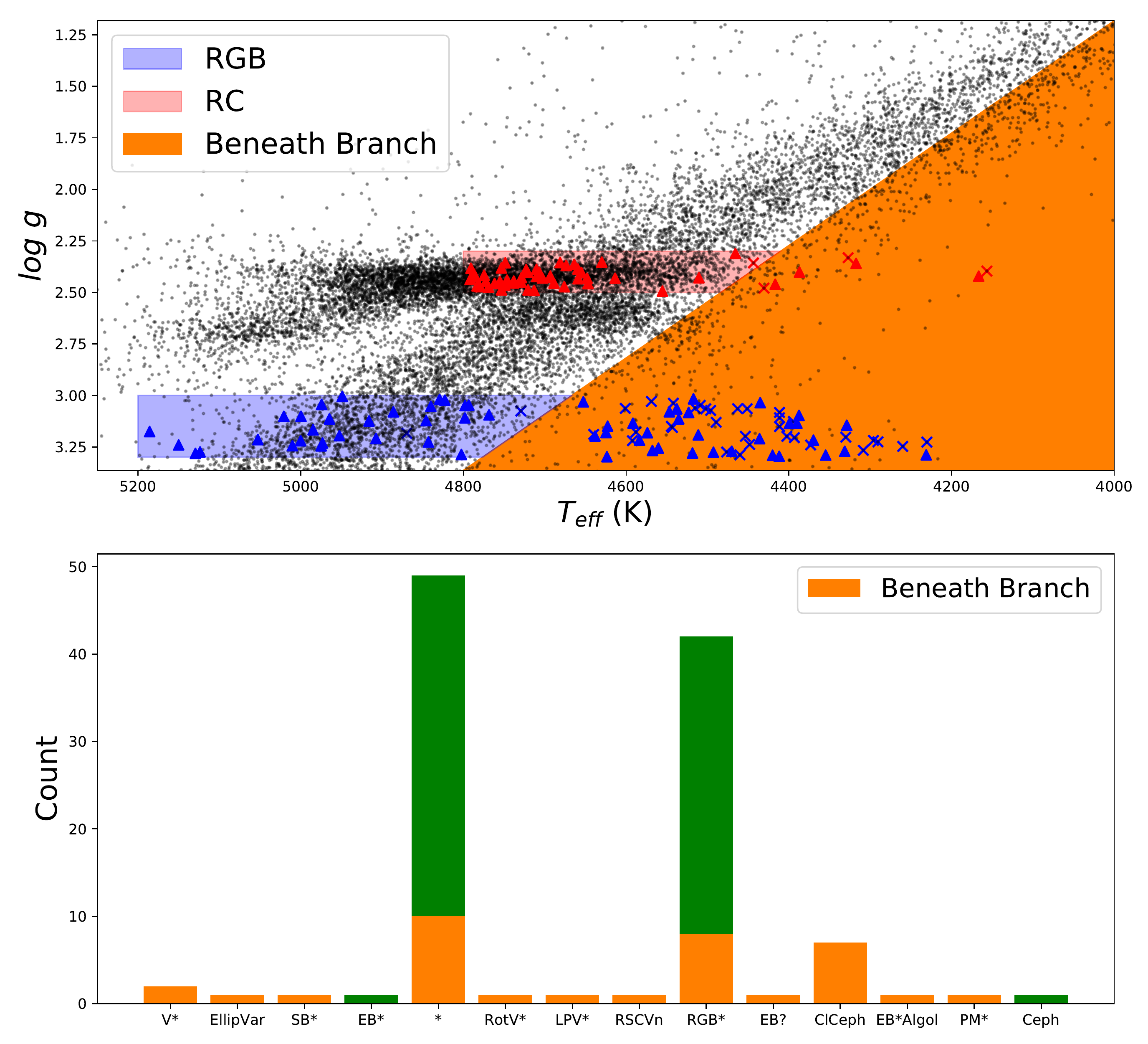}
\caption{Top: Observational HR Diagram of our Field (single) and RVvar/Binary (binary) samples using APOGEE derived $\log g$ and effective temperature. The colored rectangles and data points are our applied filters for RGB (blue) and RC (red) evolutionary states. Triangle points are stars classified in Simbad \red{(110)} and x points are stars not classified in Simbad \red{(41)}. The orange region to the right of the giant branch is labeled as {beneath branch} 
to highlight potential contaminating sources. Bottom: Bar chart of Simbad stellar classification matches to our giants. For quality control we remove classifications outside of star, red giant branch (RGB*) and eclipsing binary (EB*). 
}
\label{fig:hr_obs}
\end{figure*}

Figure~\ref{fig:hr_obs} depicts the sampling of giants from these catalogs in the Hertzsprung-Russell diagram, as well as stellar classification distributions from Simbad. We note that some giants in our sample lie to the right of (or below) the nominal red giant branch, which is unexpected. The Simbad classifications reveal that virtually all of the stars with unusual classifications (e.g., RS CVn, Ellipsoidal Variables, etc.) are in this region. Therefore, to help prevent contamination of our samples we dropped anything that had any kind of unusual Simbad classification, i.e., not ``RGB*", ``EB*", or simply ``*". We did examine the effect of removing all stars below the giant branch; our major results are unchanged, albeit with lower statistical significance. Our final cut resulted in the removal of 21 stars from our samples.

While $T_{\rm eff}$ and $\log g$ of the stars measured by APOGEE are well calibrated and thus appropriate for characterization, the vast majority of stars do not have rotational velocities (\vsini) measured by the ASPCAP pipeline. For this reason, we follow the procedure described in \citet{2015ApJ...807...82T} to determine the amount of additional broadening needed to bring the best-fit synthetic spectrum into agreement with the observation. 
Because the APOGEE spectrum is divided across three detectors, our final derived broadening value is calculated by averaging the derived $v\sin i$ for each detector. If there is a $v\sin i$ value directly reported by APOGEE we include it in the calculation of the mean. \red{As our goal in this study is to relate activity measures to stellar rotation, we sought to avoid the sample being dominated by the vast majority of very slowly rotating giants, therefore stars with no measurable $v\sin i$ (or whose uncertainty in $v\sin i$ is larger than the measurement itself) have been discarded for the purposes of this study sample. {\bf This cut removed $97\%$ of the Field sample, which is consistent with previously determined estimates of slow rotators on the giant branch \citep{1999A&AS..139..433D,2011ApJ...732...39C}.}}



\subsection{2MASS and GALEX Photometry}

\begin{figure}[!ht]
\centering
\includegraphics[width=\linewidth]{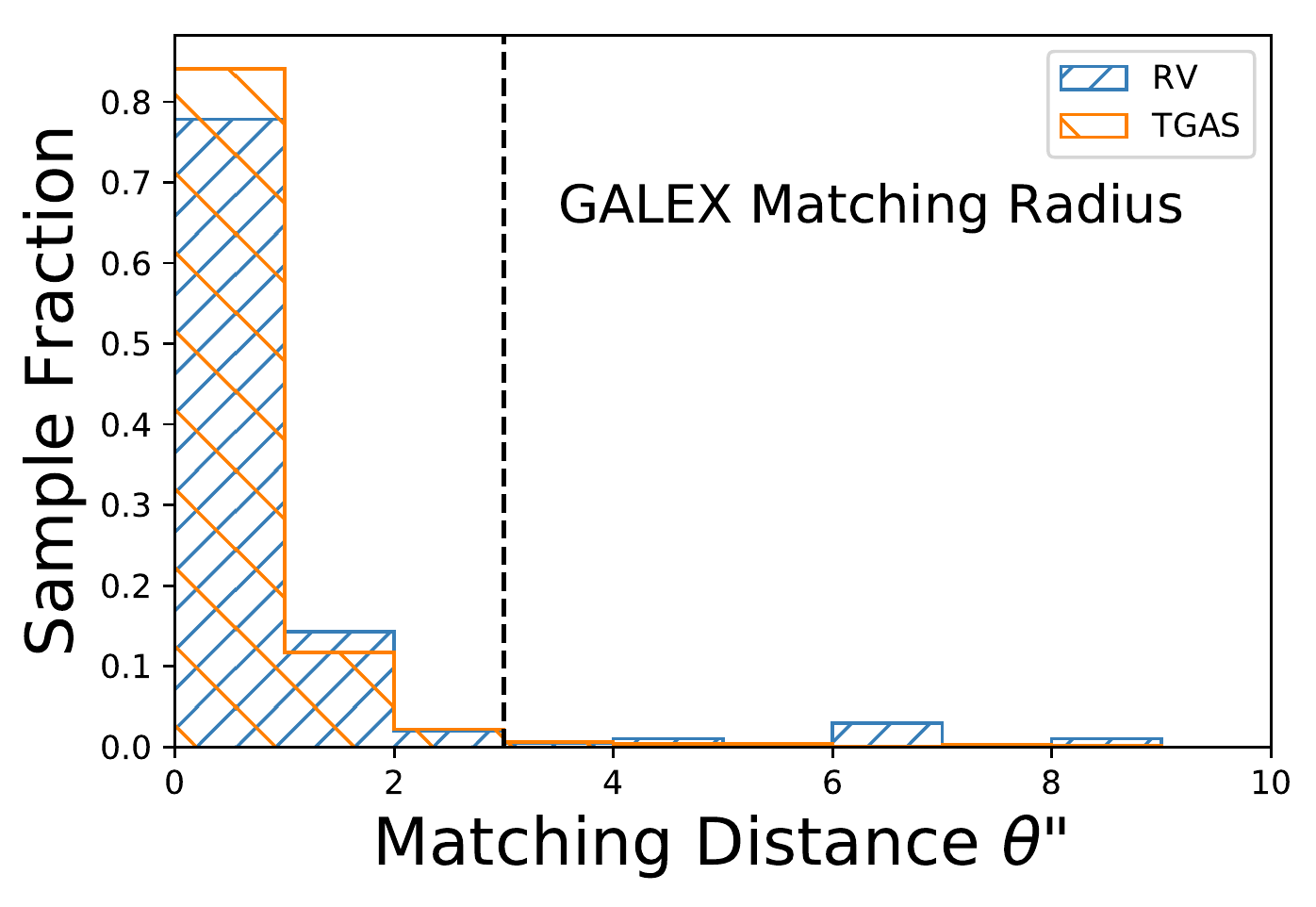}
\caption{Distribution of the closest GALEX matches during radial query in bins of 1 arsecond. The dashed vertical line marks the determined matching radius limit.}
\label{fig:galex_matching_distribution}
\end{figure}

Our giants were matched to both the {\it 2MASS\/} and {\it GALEX\/} surveys to record $J$, $K_S$, and $NUV$ magnitudes. Because APOGEE uses {\it 2MASS\/} names for its targets, our matching to {\it 2MASS\/} for the $J$ and $K_S$ photometry was \blu{unambiguous}.  
To record $NUV$ magnitudes from {\it GALEX\/}, we crossmatched the {\it 2MASS\/} positions to the nearest {\it GALEX\/} star, excluding cases where the $NUV$ value was reported as $-999$ or {\tt null}. An examination of the closest match distance distribution (Figure~\ref{fig:galex_matching_distribution}) suggested a final matching radius of $3''$ to ensure a recovery of well over $90\%$ and to limit spurious matching. In combination with our APOGEE constraints, this photometric crossmatching netted \red{78 giants (32 RGB, 46 RC)} from the Field sample and \red{63 giants (54 RGB, 9 RC)} from the RVvar/Binary sample. An overlap between the Field and RVvar/Binary samples of \red{8} giants resulted in 133 unique giants for subsequent data analysis.

\subsection{Defining UV Excess}
\label{uv_excess}
\begin{figure*}[!ht]
\centering
\includegraphics[width=.6\paperwidth]{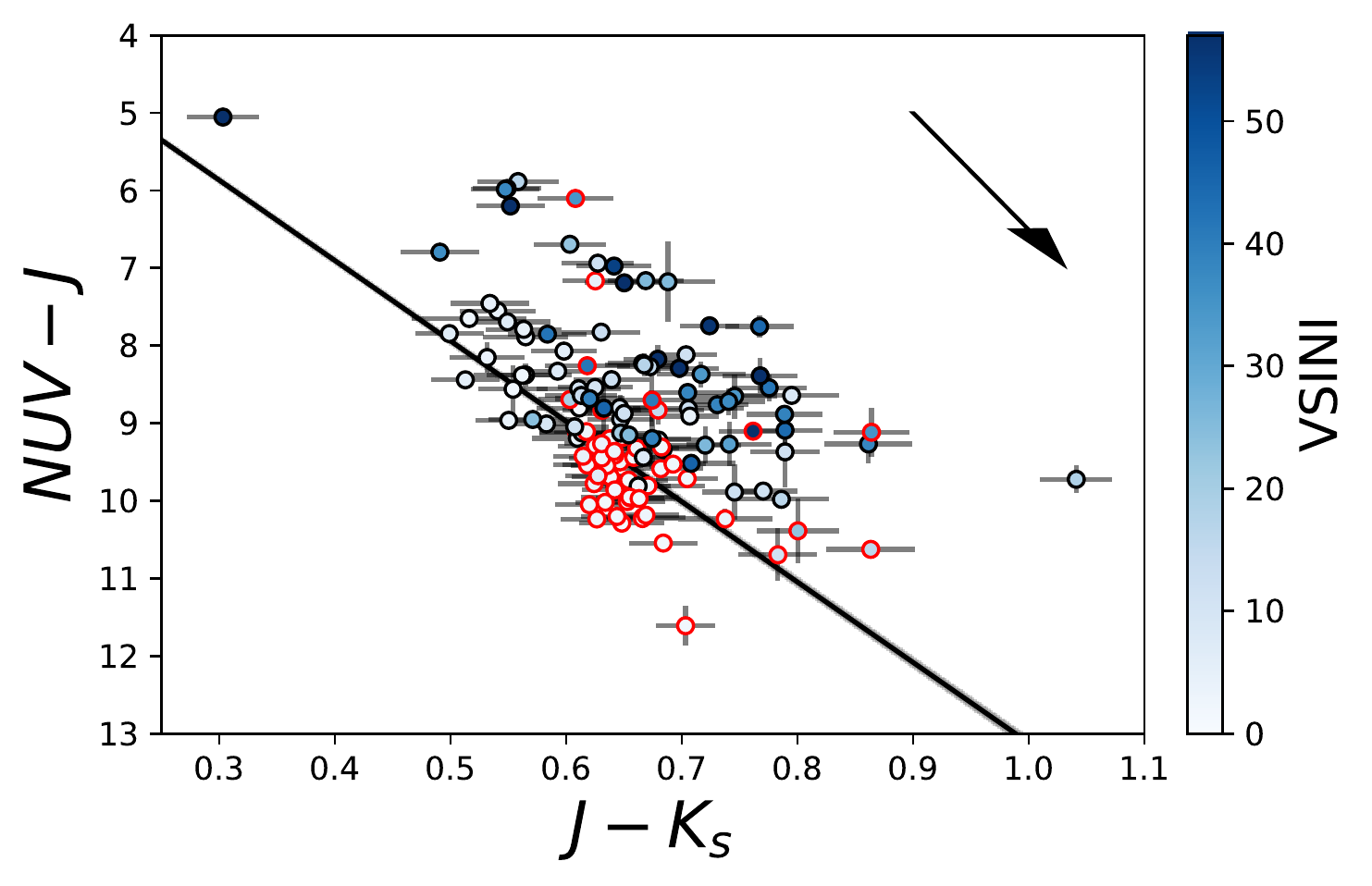}
\caption{Color-color diagram of the giants in our study sample, with {\bf blue color scale} representing $v\sin i$ {\bf in units of km/s}. Black symbol edges are for RGB stars 
and red edges are for RC stars. The solid black line represents the reference relationship relative to which $NUV$ excess is measured. The black arrow in the top right corner represents the reddening vector of the color space and each data point is corrected for reddening based on position in the Pan-STARRS 3D dust map.
}
\label{fig:color_color}
\end{figure*}

\citet{2010AJ....139.1338F} found that the colors they calculated via Kurucz atmosphere models suggested characteric stellar loci in $NUV$-{\it 2MASS\/} color spaces. To precisely calibrate empirical loci of $NUV$ excess relative to bare photospheres in the poorly tested ultraviolet regime the authors used a mixture of dwarfs and giants in the field. We utilize the \citet{2010AJ....139.1338F} locus calibrated by the Taurus field in the $NUV-J$ versus $J-K_S$ color space as the floor of excess ultraviolet activity ($NUV$ floor): 
\begin{equation}
    NUV-J=(10.36 \pm 0.07)(J-K_S) + (2.76 \pm 0.04)
    \label{eq:taurus_locus}
\end{equation}
and defined $NUV$ excess in units of magnitude as the vertical displacement from the NUV floor (Figure~\ref{fig:color_color}). Note that, in the usual sense of magnitudes, a more negative excess corresponds to a stronger excess. 

To account for extinction we dereddened our stars using the 3D dustmap from the Pan-STARRS~1 and {\it 2MASS\/} surveys \citep{2018MNRAS.478..651G}. We queried this dust map for selective extinction values using the APOGEE sky positions and {\it Gaia\/} parallaxes of our giants. Using the extinction coefficients for {\it 2MASS\/} and {\it GALEX\/} passbands reported by \citet{2013MNRAS.430.2188Y}, we perform a color transformation on the selective extinction to find excess reddening in both axes of the $NUV-J$ versus $J-K_S$ color space to define a reddening vector, where the reddening $E(B-V)$ is obtained from the Pan-STARRS reddening map: 
\begin{equation}
    E(NUV-J) = 6.33E(B-V) \label{eq:nuv_J_excess}
\end{equation}
and
\begin{equation}
    E(J-K_S) = 0.42E(B-V). \label{eq:J_Ks_excess} 
\end{equation}

\red{As in prior work \citep{Findeisen2011}, $NUV$ excess so defined is a continuous quantity, ranging from stars with large excess (large {\it negative} displacements relative to the locus) to those that are consistent with zero excess. Because of observational noise in the $NUV-J$ colors, some stars with zero true excess may scatter to non-physical $NUV-J$ values (i.e., positive displacements relative to the excess); indeed, a small number of stars in Figure~\ref{fig:color_color} appear slightly below (positive displacement) our adopted locus. {\bf We identify 25 stars} that are significantly above ($1\sigma$ negative displacement) our adopted locus using the values provided in Table~\ref{tab:data}. In the analysis that follows, we include the full sample so as to not bias our activity relations against stars that have $NUV-J$ excesses statistically consistent with zero}.

\red{M-dwarf companions to the red giants in our sample could potentially be sources of $NUV$ excess contamination. To assess this, we again use the empirical results of \citet{Findeisen2011}, who found that active K and M dwarfs of similar $J-K_S$ color to our giants can exhibit $NUV-J$ excesses as large as $-3.5$ mag. If we assume that a typical giant in our sample has such an active companion, we can calculate what would be the observed $NUV-J$ excess. The result is that a very active dwarf companion would result in an apparent $NUV-J$ excess of only about $-0.01$ mag, and cannot explain the observed excesses that are as large as $-3$ mag in some cases (see Figure~\ref{fig:color_color}).} 

\begin{figure*}[!ht]
\centering
\includegraphics[width=.6\paperwidth]{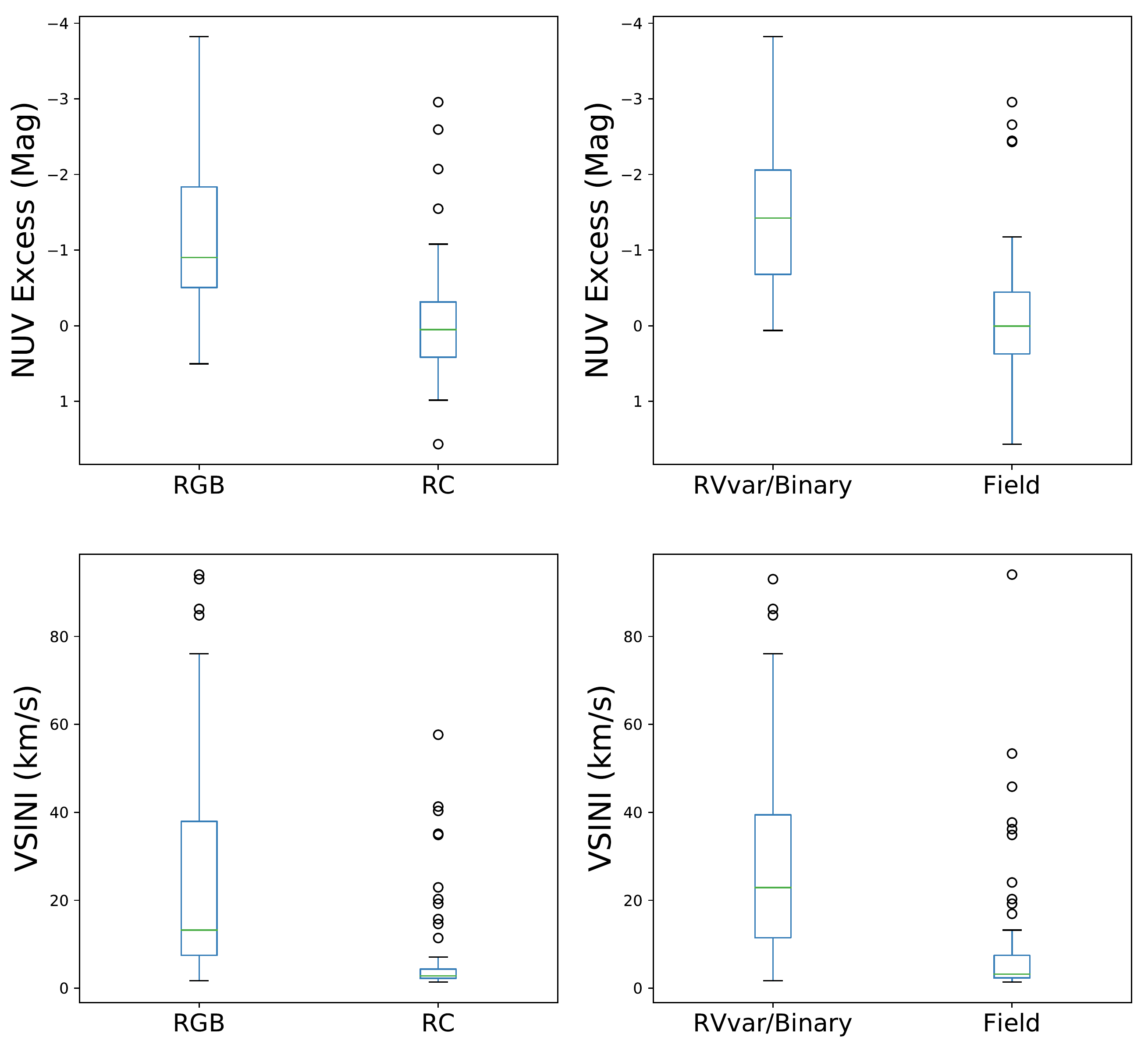}
\caption{\red{$NUV$ excess boxplot distributions between evolutionary states and subsamples.} {\bf Boxes represent the inter-quartile range, with the median indicated. Bars represent $\pm$1.5 times the inter-quartile range, or the full spread of the data, whichever is smaller.} Outliers are represented as open circles. }
\label{fig:box_plot}
\end{figure*}

The bisection of our $NUV$ excess data by the field-star locus shows giants with high rotation measures to be primarily in regions of high $NUV$ excess, supporting the expectation that stellar activity is related to rotation. $NUV$ excess distributions of RGB and RC evolutionary states show that the RC stars are more tightly concentrated around low $NUV$ excess (Figure~\ref{fig:box_plot}); 
this is consistent with expectations as our RC $T_{\rm eff}$ and $\log g$ cutoffs select stars of lower mass compared to our RGB temperature cutoffs, and lower mass giants generally rotate more slowly. Also, RC stars are more evolved and have had additional time to spin down.
{\bf In addition, these stars are less likely to have close companions that would tidally drive rapid rotation, as evidenced by their deficit in the RVvar/Binary sample.}

\begin{figure}[!ht]
\centering
\includegraphics[width=\linewidth]{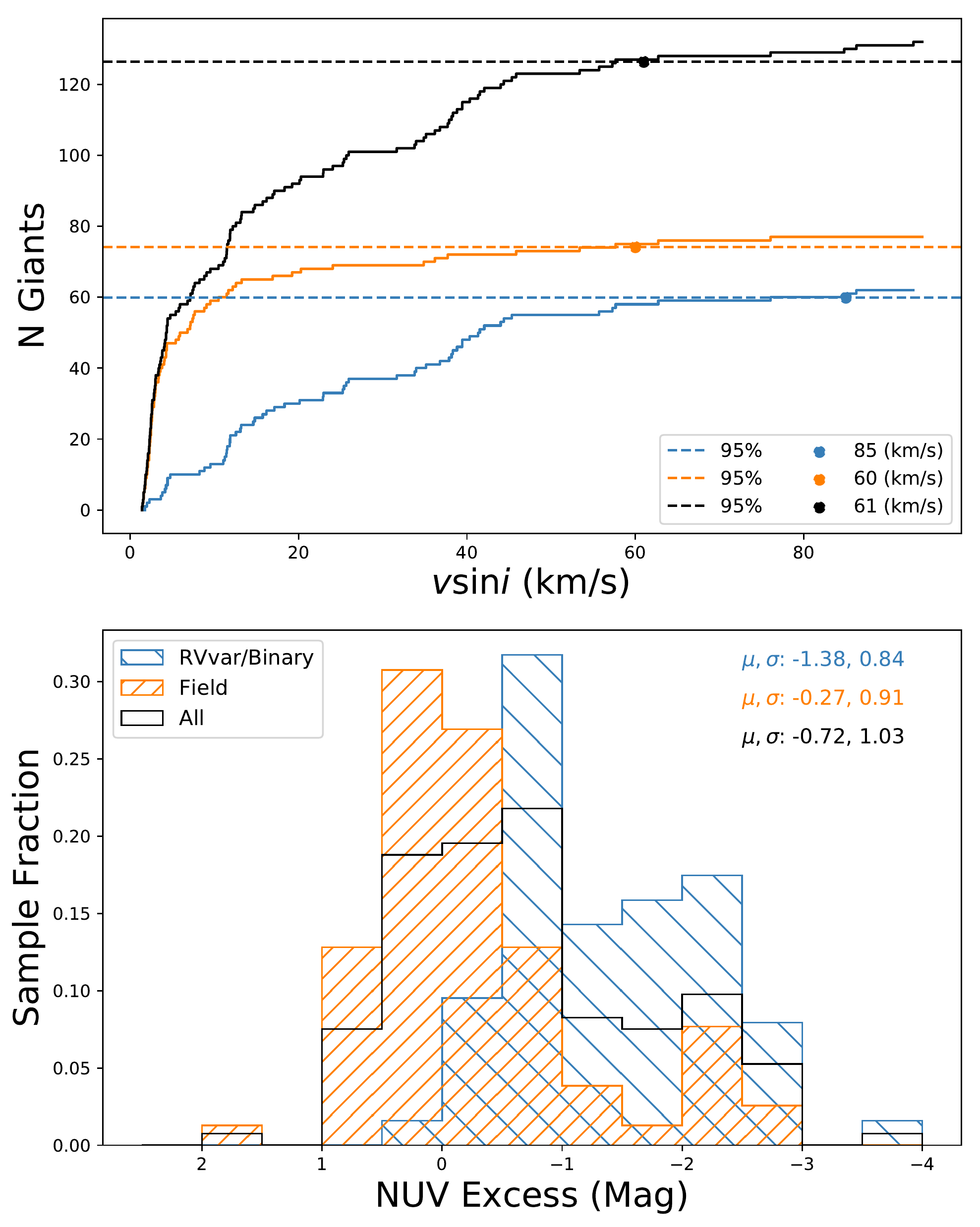}
\caption{Top: Cumulative distribution of the \red{133} giants used to calibrate our $NUV$ excess relations. Black, orange and blue colors represent all giants, Field, and RVvar/Binary, respectively; the dots and dashed lines mark the $95^{th}$ percentile for each sample. Bottom: Histogram of $NUV$ excess separated by sample membership in bins of 0.5~mag.
}
\label{fig:distributions}
\end{figure}

\blu{Finally, Figure~\ref{fig:distributions} shows the distributions of $NUV$ excess for the Field and RVvar/Binary samples separately, and for the full sample combined. Not surprisingly, the RVvar/Binary sample is significantly skewed to larger (more negative) $NUV$ excess, likely due to the fact that the red giants with binary companions are more often found to be rotating rapidly. This agrees with the interpretation that $NUV$ excess is linked to rotation, as we now discuss.}

\section{Results: Empirical Rotation-Activity Relation for Red Giants}
\label{results}
In this section we present the results of the empirical relationships between stellar rotation ($v \sin i$, rotation period, and Rossby number) and chromospheric activity as measured by $NUV$ excess. We present the results for the nominal binary sample, the nominally single sample, and in combination. 

\begin{figure*}[!ht]
\centering
\includegraphics[width=.65\paperwidth]{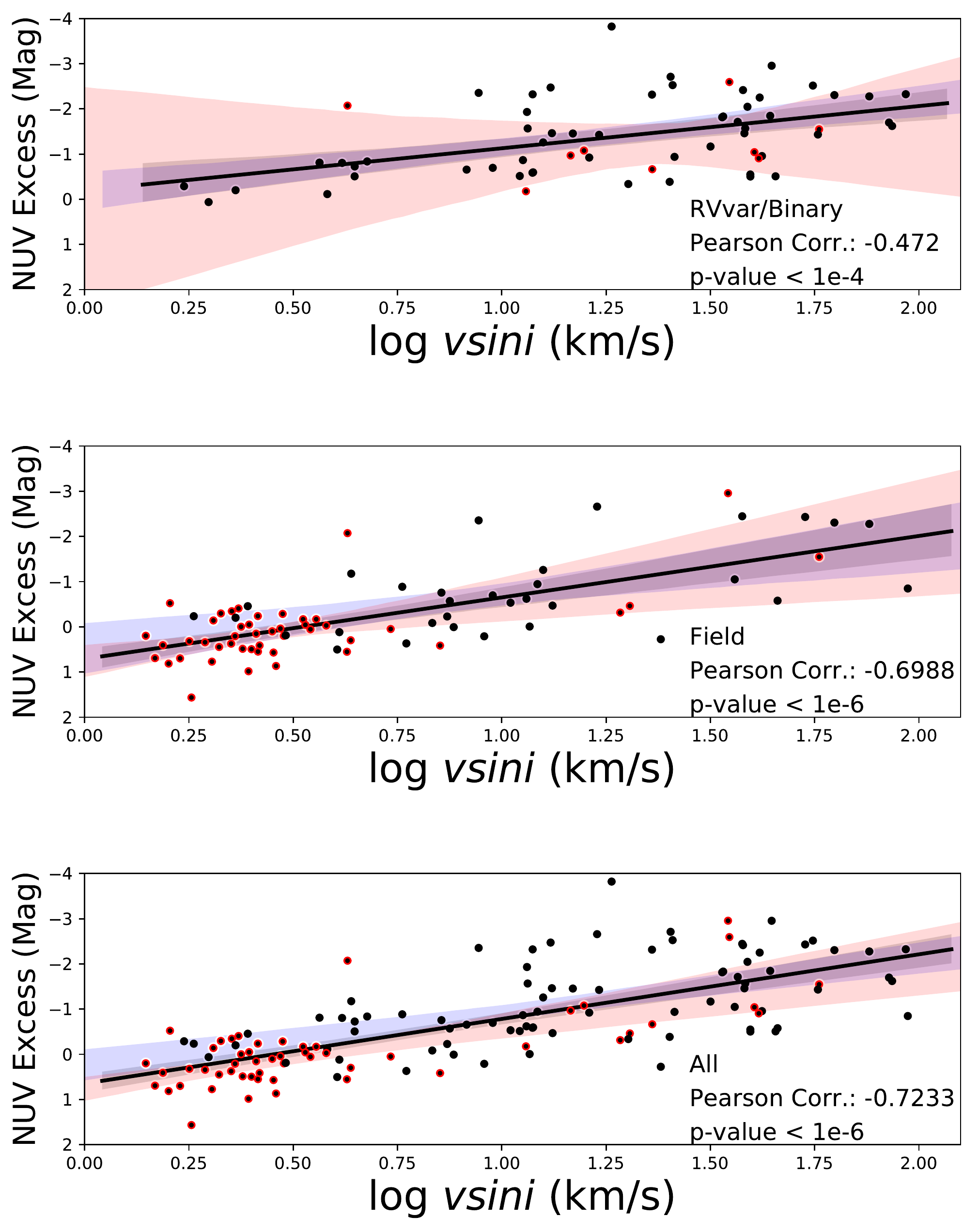}
\caption{$NUV$ excess versus log $v\sin i$ for the  RVvar/Binary (top), Field (middle) and combined samples (bottom)  
Best fit lines for RGB (blue) and RC (red) evolutionary states were fit, with the black lines depicting the best fit for the combined sample. \red{Each subplot reports the sample name along with the Pearson correlation coefficient and significance. Red edges identify stars in the RC phase.} 
}
\label{fig:nuv_vsini}
\end{figure*}

In Figure~\ref{fig:nuv_vsini}, we plot $NUV$ excess against $v\sin i$ for our red giant samples, and we apply first-order fits using linear regression. To visualize potential dependencies on  evolutionary state, we plot fits for RGB and RC stars separately. 

\blu{The most straightforward, and important, finding is that to first order there is a highly statistically significant correlation (false-alarm probability less than $10^{-4}$) between $NUV$ excess and $v\sin i$ for both the Field and RVvar/Binary samples individually and in combination. Moreover, a single fit relation (Equation~\ref{eq:excess_equation}) using log vsini adequately fits the full combined sample, with a Pearson's \red{$r=-0.72$} ($p$-value $<10^{-6}$) as well as a Kendall's \red{$\tau=-0.54$} at similarly high statistical significance (p-value $< 10^{-6}$):
\begin{equation}
    y=(-1.43 \pm 0.12)x + (0.647 \mp 0.131) 
    \label{eq:excess_equation}
\end{equation}
where $y$ is the $NUV$ excess and $x$ is $\log v\sin i$.
{\bf The relations for the RVvar/Binary, Field, and full sample best fit lines take the form of $y=(-0.934 \pm 0.227)x + (-0.194 \mp 0.305)$, $y=(-1.36 \pm 0.162)x + (0.711 \mp 0.14)$, and $y=(-1.43 \pm 0.12)x + (0.647 \mp 0.131)$, respectively.}
} 

\begin{table*}
    \centering
        \def\arraystretch{1.5}
        \hspace*{-2.25cm}\begin{tabular}{lrrrrrrrrr}
        \toprule
        {} &  NUV exess &  $v \sin i$ &  $P_{rot}/ \sin i$ &     $T_{\rm eff}$ &  $\log g$ &  mass &  radius &   [M/H] &   [C/N] \\
        \midrule
        NUV exess &       1.00 &  -0.41 &    0.33 &  0.43 & -0.08 &  0.25 &    0.21 &  0.03 & -0.00 \\
        $v \sin i$     &      -0.41 &   1.00 &   -0.74 & -0.40 &  0.03 & -0.27 &   -0.29 & -0.02 &  0.10 \\
        $P_{rot}/ \sin i$    &       0.33 &  -0.74 &    1.00 &  0.29 & -0.20 &  0.52 &    0.56 &  0.48 & -0.21 \\
        $T_{\rm eff}$         &       0.43 &  -0.40 &    0.29 &  1.00 & -0.07 &  0.35 &    0.29 & -0.10 & -0.12 \\
        $\log g$      &      -0.08 &   0.03 &   -0.20 & -0.07 &  1.00 & -0.07 &   -0.23 &  0.04 &  0.02 \\
        mass      &       0.25 &  -0.27 &    0.52 &  0.35 & -0.07 &  1.00 &    0.83 &  0.35 &  0.03 \\
        radius    &       0.21 &  -0.29 &    0.56 &  0.29 & -0.23 &  0.83 &    1.00 &  0.37 &  0.06 \\
        {[M/H]}       &       0.03 &  -0.02 &    0.48 & -0.10 &  0.04 &  0.35 &    0.37 &  1.00 &  0.09 \\
        {[C/N]}       &      -0.00 &   0.10 &   -0.21 & -0.12 &  0.02 &  0.03 &    0.06 &  0.09 &  1.00 \\
        \bottomrule
        \end{tabular}
    \caption{Kendall Tau correllation matrix for the 80 RGB stars in our dataset, of which 31 have determined radii. }
\end{table*}

\begin{table*}
    \centering
        \def\arraystretch{1.5}
        \hspace*{-2.25cm}\begin{tabular}{lrrrrrrrrr}
        \toprule
        {} &  NUV exess &  $v \sin i$ &  $P_{rot}/ \sin i$ &     $T_{\rm eff}$ &  $\log g$ &  mass &  radius &   [M/H] &   [C/N] \\
        \midrule
        NUV exess &       1.00 &  -0.42 &    0.21 &  0.12 &  0.11 &  0.03 &   -0.03 &  0.28 & -0.03 \\
        $v \sin i$     &      -0.42 &   1.00 &   -0.85 & -0.07 &  0.07 & -0.00 &   -0.00 & -0.14 &  0.17 \\
        $P_{rot}/ \sin i$    &       0.21 &  -0.85 &    1.00 & -0.11 & -0.18 &  0.14 &    0.15 &  0.09 & -0.32 \\
        $T_{\rm eff}$       &       0.12 &  -0.07 &   -0.11 &  1.00 &  0.17 & -0.07 &   -0.21 & -0.20 & -0.10 \\
        $\log g$      &       0.11 &   0.07 &   -0.18 &  0.17 &  1.00 &  0.33 &   -0.02 &  0.21 & -0.04 \\
        mass      &       0.03 &  -0.00 &    0.14 & -0.07 &  0.33 &  1.00 &    0.65 &  0.11 & -0.08 \\
        radius    &      -0.03 &  -0.00 &    0.15 & -0.21 & -0.02 &  0.65 &    1.00 &  0.06 &  0.03 \\
        {[M/H]}      &       0.28 &  -0.14 &    0.09 & -0.20 &  0.21 &  0.11 &    0.06 &  1.00 & -0.11 \\
        {[C/N]}       &      -0.03 &   0.17 &   -0.32 & -0.10 & -0.04 & -0.08 &    0.03 & -0.11 &  1.00 \\
        \bottomrule
        \end{tabular}
   \caption{Kendall Tau correllation matrix for the 53 RC stars in our dataset, of which 46 have determined radii. }
\end{table*}

\begin{table*}
    \centering
        \def\arraystretch{1.5}
        \hspace*{-2.25cm}\begin{tabular}{lrrrrrrrrr}
        \toprule
        {} &  NUV exess &  $v \sin i$ &  $P_{rot}/ \sin i$ &     $T_{\rm eff}$ &  $\log g$ &  mass &  radius &   [M/H] &   [C/N] \\
        \midrule
        NUV exess &       1.00 &  -0.54 &    0.42 &  0.33 & -0.32 & -0.09 &    0.10 &  0.15 &  0.07 \\
        $v \sin i$     &      -0.54 &   1.00 &   -0.87 & -0.30 &  0.33 &  0.15 &   -0.14 & -0.13 &  0.01 \\
        $P_{rot}/ \sin i$    &       0.42 &  -0.87 &    1.00 & -0.08 & -0.46 & -0.05 &    0.27 &  0.30 & -0.01 \\
        $T_{\rm eff}$         &       0.33 &  -0.30 &   -0.08 &  1.00 & -0.05 &  0.28 &    0.06 & -0.09 & -0.11 \\
        $\log g$      &      -0.32 &   0.33 &   -0.46 & -0.05 &  1.00 &  0.40 &   -0.16 & -0.05 & -0.09 \\
        mass      &      -0.09 &   0.15 &   -0.05 &  0.28 &  0.40 &  1.00 &    0.44 &  0.11 & -0.12 \\
        radius    &       0.10 &  -0.14 &    0.27 &  0.06 & -0.16 &  0.44 &    1.00 &  0.21 &  0.01 \\
        {[M/H]}       &       0.15 &  -0.13 &    0.30 & -0.09 & -0.05 &  0.11 &    0.21 &  1.00 &  0.04 \\
        {[C/N]}       &       0.07 &   0.01 &   -0.01 & -0.11 & -0.09 & -0.12 &    0.01 &  0.04 &  1.00 \\
        \bottomrule
        \end{tabular}
 \caption{Kendall Tau correllation matrix for the 133 giants in our dataset, of which 77 have determined radii. }
\end{table*}

We also solve for three Kendall's $\tau$ correlation matrices, one for each evolutionary state (RGB and RC) and their combination, in order to estimate the effects of multiple parameters simultaneously. Between evolutionary states we find that RGB stars are driving the strong dependency of $NUV$ excess on rotation, since RGB stars are more likely to rotate rapidly compared to RC stars. We also surprisingly find that lower temperature stars appear to have a {\it greater} $NUV$ excess when physically the correlation should be in the other direction. This may indicate an issue with 
the ASPCAP calculated temperature for broadened spectra because ASPCAP does not include a $v\sin i$ dimension in its model grid for giants. Fortunately, these residual correlations are mitigated in the combined sample, such that the rotational parameters $v\sin i$ and $P_{\rm rot}$ dominate over these other correlations. Finally, we find weak to moderate correlation values between $NUV$ excess and both $\log g$ and [M/H], with stellar mass, radius, and [C/N] ratio appearing negligible in their contributions. Despite other measurable dependencies it is clear in every matrix that $v\sin i$ (and $P_{\rm rot}$; see below) are the dominant correlates of $NUV$ excess.

For completeness, we do note that there may be a modest difference in the $NUV$ excess versus $v\sin i$ relation for the RVvar/Binary (putative binary) and Field samples. In particular, the RVvar/Binary relation is somewhat shallower {and has a slightly higher excess at very slow rotation but converges to the Field relation at faster rotation.} 
We also note that the binary sample has a more symmetric $NUV$ excess distribution compared to that for the singles (see Figure~\ref{fig:distributions}b). Thus we cannot definitively rule out that the relation between $NUV$ excess and $v\sin i$ could in fact be somewhat different among the red giants with binary companions compared to those likely to be single. 
Next, in order to estimate rotation periods for our sample we first solve for stellar radii:
\begin{equation}
    R=\frac{\theta _{dia} \times d_{gaia}}{2}
    \label{eq:radius_equation}
\end{equation}
where $\theta _{dia}$ is the reported angular diameters in \citet{2017AJ....154..259S} and $ d_{gaia}$ is the distance derived from Gaia parallax. Using this radius we solve for rotation period:
\begin{equation}
    \frac{P_{\rm rot}}{\sin i}=\frac{2\pi R}{v\sin i}
    \label{eq:psini_equation}
\end{equation}
where the rotation period measure ($P_{\rm rot}/\sin i$) is an upper limit on the true rotation period ($P_{\rm rot}$) because of the indeterminate $\sin i$ that is inherited from the $v\sin i$ measurements. We successfully obtained $P_{\rm rot}/\sin i$ measures for \red{77} of our red giant sample. 
Only 1 of the \red{77} stars with $P_{\rm rot}/\sin i$ measures is uniquely in the binary sample, effectively limiting the $P_{\rm rot}/\sin i$ sample to be inclusive of only Field stars. 

One standard measure that uses $P_{\rm rot}$ to quantify magnetic activity is the Rossby number ($R_0$), which is defined as:
\begin{equation}
    R_0=P_{\rm rot}/\tau_{\rm turnover}
    \label{eq:ross_equation}
\end{equation}
where $\tau _{turnover}$ is the convective overturn timescale.
{\bf Because evolved stars exhibit qualitatively different interior properties compared to main-sequence stars,} we estimated the convective overturn timescale of our giants using an empirical relation between $T_{\rm eff}$ and $\tau _{turnover}$ calibrated {\bf specifically for giant stars} \citep{1987ApJ...316..377B}.

\begin{figure*}[!ht]
\centering
\includegraphics[width=0.5\paperwidth]{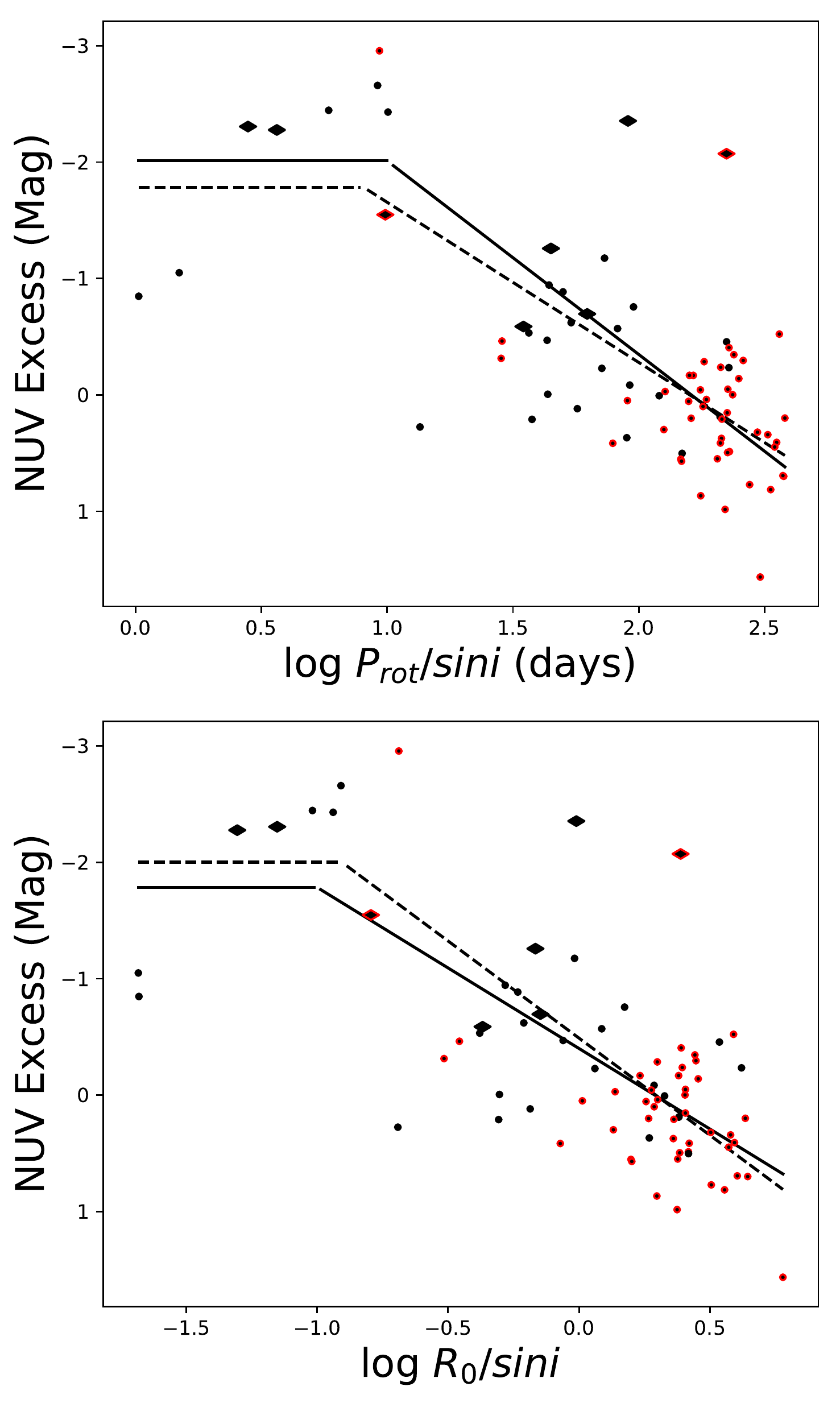}
\caption{$NUV$ excess versus $\log(P_{\rm rot}/\sin i)$ (top) and Rossby number (bottom), for giants in our sample with calculated stellar radii. \red{Solid lines represent a fit at a fixed saturation period, and dashed lines represent a fit at a fixed Rossby number threshold, based on the corresponding M-dwarf saturation threshold and estimated convective overturn timescale.  Black symbol edges are for RGB, red edges are for RC and diamonds indicate binary sample membership.}}
\label{fig:nuv_period}
\end{figure*}

{\bf 
Plotting $NUV$ excess against $P_{\rm rot}/\sin i$ (Figure~\ref{fig:nuv_period}a) we noticed an apparent saturation region for $P_{\rm rot}/\sin i < 10$~d. This behavior has been previously observed {\bf in X-rays} in M~dwarf stars, {\bf and tentatively identified in $NUV$ as well} \citep[see, e.g.,][]{2016MNRAS.463.1844S}. The same trend is noticed in Rossby number (Figure~\ref{fig:nuv_period}b).

The saturated region was fit with a flat line and the non-saturated region was fit with a sloped line. 
We adopted the same thresholds for the saturated region, in terms of $P_{\rm rot}$ and Rossby number, seen in M~dwarfs \citep[$\log P_{\rm rot}=1$ and $\log R_0=-1.0$;][]{2016MNRAS.463.1844S}. 
The result of fitting at both thresholds are shown in Figure~\ref{fig:nuv_period}.
}

\red{The general similarity with the M~dwarf behavior may suggest similar dynamo mechanisms between the two kinds of stars (see Section~\ref{sec:mdwarfs})}. 
We also note that there is no significant difference between the RGB and RC phases which, like $v\sin i$, suggests only a weak dependence on evolutionary state.

\section{Discussion}
\label{discussion}

\blu{In this section we discuss how our newly determined relations for $NUV$ excess as functions of $v\sin i$, $P_{\rm rot}/\sin i$, and $R_0$ compare to those for similarly cool stars that are still on the main sequence (i.e., M~dwarfs). 
We discuss the implications of the similarities in these relations for what may be underlying similarities in the physical processes driving activity in red giants and cool dwarfs. 
Finally, we briefly present a novel application of our new rotation-activity relations.}

\subsection{Comparison to M-dwarf Activity}\label{sec:mdwarfs}
\subsubsection{General rotation-activity relation}
To examine the relationship between rotation and stellar activity for low-mass main-sequence stars, \cite{2016MNRAS.463.1844S} compared {\it GALEX\/} $NUV$ emission to photometrically derived rotation periods for a sample of 32 M~dwarfs cross-matched between the Superblink catalog and the K2 mission. We use their reported {\it 2MASS\/} and {\it GALEX\/} photometry for this M-dwarf sample to estimate $NUV$ excess via our definition (Equation~\ref{eq:taurus_locus}). 
We 
also derive rotational velocity from the reported rotation periods and radii. {\bf To ensure quality, M~dwarfs with periods that were flagged as unreliable were discarded. Additionally, one M~dwarf was removed for having a reported period ($>100$~d), longer than the observation window of the data used to estimate it, leaving 18 M~dwarfs for our comparison.}

\begin{figure*}[!ht]
\centering
\includegraphics[width=0.5\paperwidth]{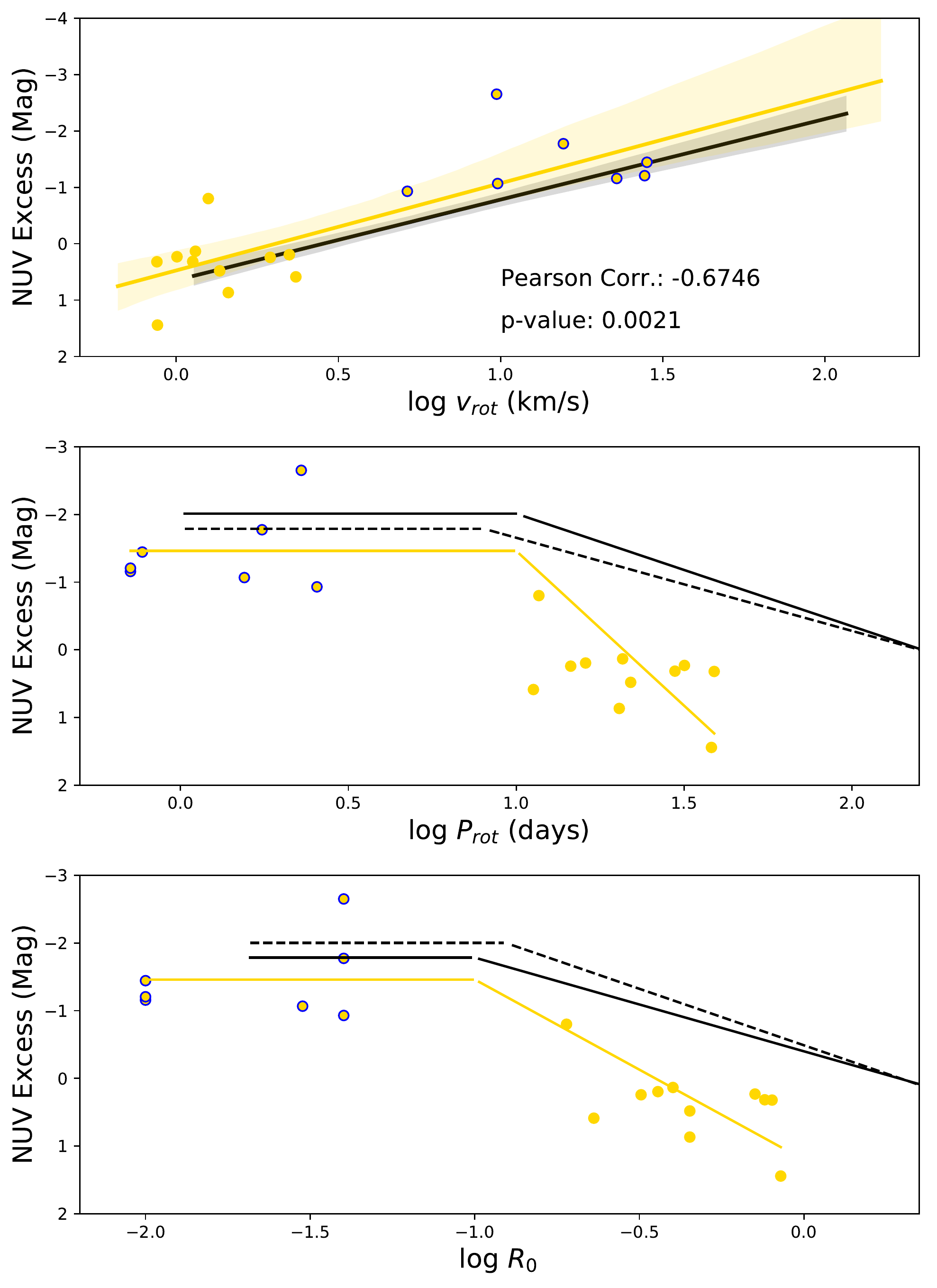}
\caption{$NUV$ excess versus $v_{rot}$ (top), $\log P_{\rm rot}$ (middle) and rossby number (bottom) for 18 M~dwarfs. \red{Gold} lines are the fit relations for the M~dwarfs. \red{For comparison, black solid and dashed lines represent the fits to the giants from Figure~\ref{fig:nuv_period}.} 
Points with blue edges represent stars with $P_{\rm rot} < 10$~d. 
}
\label{fig:mdwarfs_rot}
\end{figure*}

As we did with the red giant sample, we first consider the relationship between $NUV$ excess and $v_{\rm rot}$.
The resulting relation is shown in the top panel of Figure~\ref{fig:mdwarfs_rot} (gold line), where the best fitting linear relationship is given by
\begin{equation}
    y=(-1.547 \pm 0.301)x + (0.474 \mp 0.228) 
    \label{eq:mdwarf_vsini}
\end{equation}
where $y$ is $NUV$ excess and $x\equiv \log\ v_{\rm rot}$. 
Interestingly, this relation is nearly identical to what we found for the red giants, as shown by the black line in Figure~\ref{fig:mdwarfs_rot} (top). 

\subsubsection{Relation to Rossby Number}
\blu{
We can take an additional step to inquire whether other aspects of the M-dwarf activity behavior are also evident in red giants. In particular,}
the work of \citet[][and see references therein]{2003A&A...397..147P} has shown that \blu{the linear rotation-activity relation for late-type dwarf stars in X-rays ``saturates" for stars rotating faster than a certain critical rotation period}. This same saturation effect was later observed for M dwarfs in the $NUV$ filter \citep{2016MNRAS.463.1844S} for \blu{stars rotating faster than $P_{\rm rot} \approx 10$~d}. 

Therefore, in Figure~\ref{fig:mdwarfs_rot} (middle and bottom) we again show $NUV$ excess for the M-dwarf sample, but now plotted against $P_{\rm rot}$ and $R_0$ respectively, with a break between the linear and saturated regimes at $P_{\rm rot}=10$~d and $\log R_0 = -1$ respectively \citep[see, e.g.,][]{2003A&A...397..147P}. Comparing
to what we found for our red giant sample (Figure~\ref{fig:mdwarfs_rot}, \red{black lines}), we find 
very similar behavior, although  
the relation for the giants in the linear regime is 
somewhat shallower than for the M~dwarfs. 

\subsubsection{Supersaturation?}
\label{supersaturation}
One curious characteristic of the rotation-activity relation for M~dwarfs that has been reported is so-called ``super-saturation", whereby the most extremely fast rotating M~dwarfs have {\it decreased} activity compared to those in the saturated regime \citep[see, e.g.,][and references therein]{2011MNRAS.411.2099J}. Here we consider whether this suggested phenomenon might be present in the red giant sample as well. 

In M~dwarfs, \citet{2011MNRAS.411.2099J} has observed the super-saturation effect to occur at $P_{\rm rot} < 0.2$~d, noting that this corresponds roughly to breakup speed, such that the stellar corona could become centrifugally stripped. 
Therefore, if we adopt the same threshold for giants, we must adjust for the large difference in radius (and in mass), the supersaturation period scaling as $M^{-1/2} R^{3/2}$.
Taking an average radius of 5~R$_\odot$ and average mass of 1~M$_\odot$, the equivalent supersaturation limit for the red giants is then $\approx 4$~d. 

\begin{figure}[!ht]
\centering
\includegraphics[width=\linewidth]{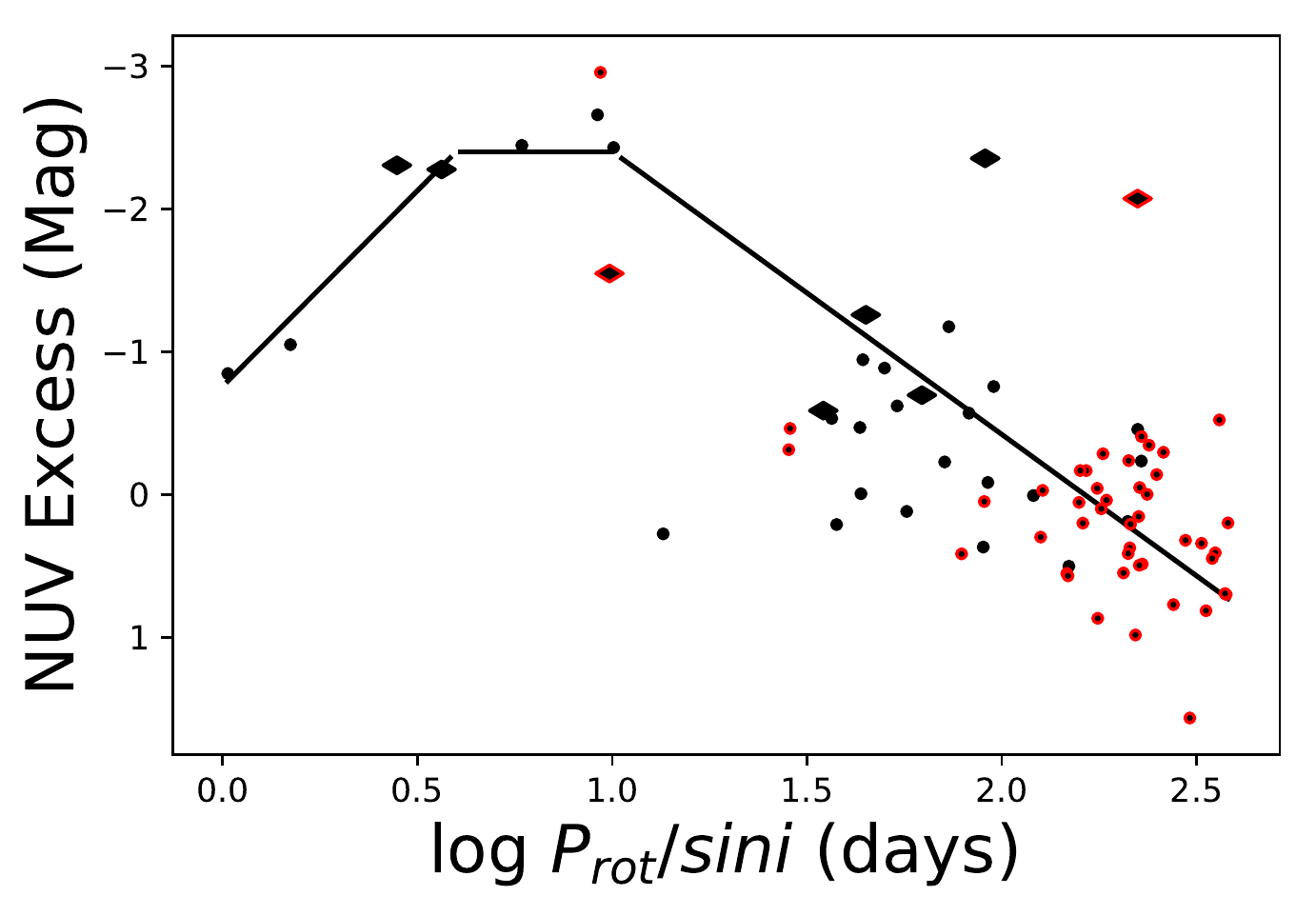}
\caption{$NUV$ excess versus $\log(P_{\rm rot}/\sin i)$, as in top panel of Figure~\ref{fig:nuv_period}, but now also including a third linear fit to the putative supersaturation regime ($\log P_{\rm rot}<0.6$~d). \red{Black symbol edges are for RGB, red edges are for RC and diamonds indicate Binary sample membership.}
}
\label{fig:supersat}
\end{figure}

We once again represent $NUV$ excess versus $\log P_{\rm rot}/\sin i$ (Figure~\ref{fig:supersat}), and fit linear and saturated regimes with a break at $P_{\rm rot} = 10$~d as before, but now include an additional linear supersaturation fit to $P_{\rm rot} < 4$~d ($\log P_{\rm rot} < 0.6$). There are only four stars in our sample that fall within the putative supersaturation regime, and perhaps only two of them that display a clear decrease in $NUV$ excess relative to the saturated level, however they do appear to be consistent with the idea.

\begin{figure}[!ht]
\centering
\includegraphics[width=\linewidth]{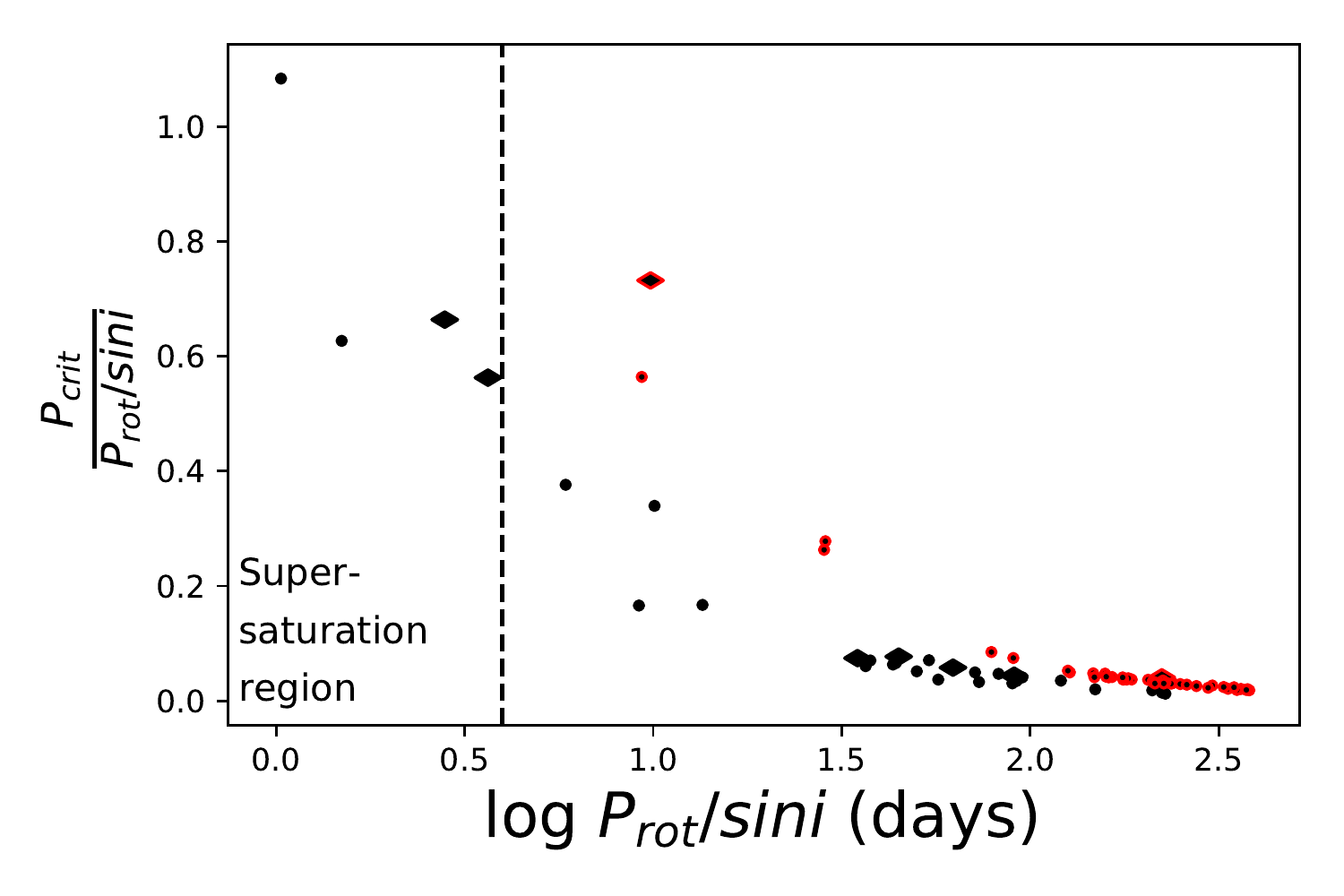}
\caption{Fraction of rotational breakup versus $\log (P_{\rm rot}/\sin i)$ for all giants with calculated $P_{\rm rot}/\sin i$ values. Vertical dashed line marks threshold for supersaturation. \red{Black symbol edges are for RGB, red edges are for RC and diamonds indicate Binary sample membership.}
}
\label{fig:pcrit}
\end{figure}

We can speculate that these supersaturated stars are due to centrifugal stripping \citep[e.g.,][]{2011MNRAS.411.2099J}, and that the stripping is proportional to the star's rotation period as a fraction of the critical rotation period, $P_{\rm crit}$, defined as the period where centrifugal force equals gravitational force \citep[see, e.g.,][]{Ceillier2017}: 
\begin{equation}
    P_{crit} \equiv \left(\frac{27\pi^2 R^3}{2GM}\right)^{1/2}
    \label{eq:pcrit}
\end{equation}
The ratio of $P_{\rm crit}$ to $P_{\rm rot}/\sin i$ for our sample is shown in Figure~\ref{fig:pcrit}, where we find that indeed the stars rotating faster than $\log P_{\rm rot}/\sin i < 0.6$ all correspond to 50--100\% of breakup speed.

\subsubsection{Summary}
One explanation for the similarity between our evolved-star and M-dwarf rotation-activity relations is the presence of rotationally driven dynamos. 
In dwarf stars, the correlation between rotation and activity is thought to be driven by photospheric and chromospheric heating driven by magnetic reconnections mediated by the underlying magnetic dynamo process. The similar relationships between rotation and UV excess identified in this work---and especially the similar linear and saturated relationships with respect to Rossby number---tends to suggest that the $NUV$ excesses we are detecting in giants are driven by an analogous magnetic dynamo. 

Despite the significant differences in surface area between our giants and the more well studied dwarfs, it is remarkable that the saturation thresholds appear so similar. Perhaps even more remarkable is the notion that our red giants may exhibit supersaturation at a comparable centrifugal rotation threshold as in M~dwarfs, suggesting yet another common mechanism ({\bf centrifugal} stripping). 
More careful investigations that fully explore potential differences in convective turnover timescale and in coronal structure, as well as substantially larger numbers of very rapidly rotating giant stars, are required to truly understand these saturation and supersaturation processes. 

The comparisons between the red giant and M~dwarf relations indicate that further work is needed to more precisely characterize the similarities and differences between them. But most fundamentally the similarities observed here are striking, which further reinforces that the trends observed in our red giant sample are not caused by contamination, viewing angle, sample selection, or some other non-intrinsic effect; $NUV$ excess is fundamentally linked to stellar rotation just as in M~dwarfs.

\subsection{Example Application: \citet{Thompson2018} red-giant/black-hole binary system}\label{sec:thompson}

\begin{figure}[!ht]
\centering
\includegraphics[width=\linewidth]{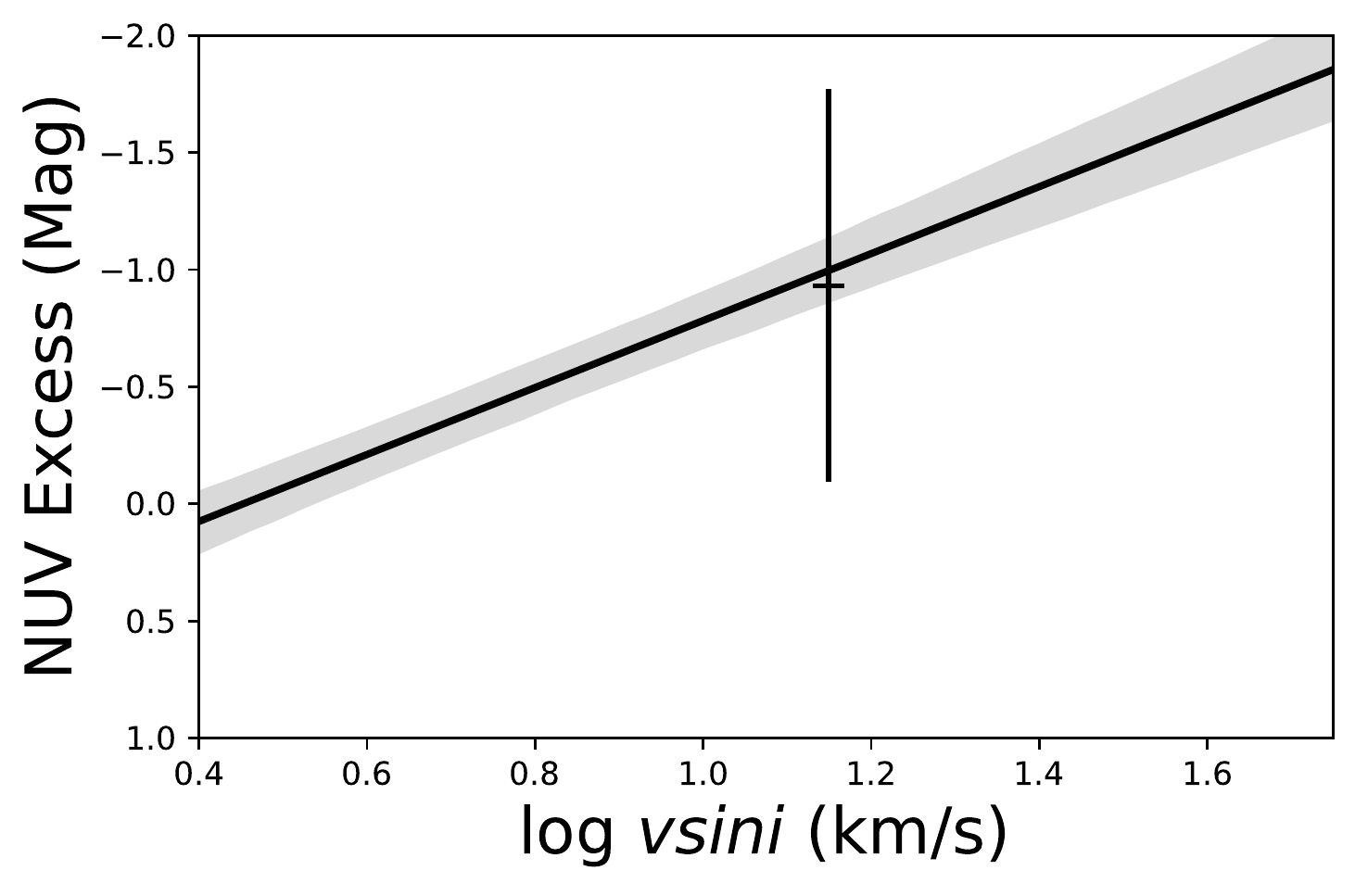}
\caption{Giant/Black Hole binary system plotted in $NUV$ excess versus $\log v\sin i$ along with our \red{fitted rotational NUV excess relation}. Plot shows the ultraviolet emission from the system is consistent with expectations from the giant. 
}
\label{fig:bh_nuv_vsini}
\end{figure}

\blu{ Recent work by \citet{Thompson2018} has identified an interesting binary system that likely includes a rapidly rotating red giant and a stellar mass black hole. They were interested in whether the black hole is actively accreting, and therefore looked for excess high energy radiation from the system. They do not detect an X-ray flux with {\it Swift}, but do detect a small $NUV$ excess with {\it GALEX\/} for this system. They use work by \citet{Bai2018} to tentatively suggest that this system has an $NUV$ excess consistent with the distribution of values in red giants, and therefore is unlikely to have significant ongoing accretion. 

With our new relation, we can show that this system has a {\it GALEX\/} $NUV$ excess consistent with \red{ what would be expected for the red giant component at its measured rotation velocity ($\sim$14~\kms, see Figure~\ref{fig:bh_nuv_vsini})} and we can thus confidently conclude that there is no detectable accretion by this stellar mass black hole.}

\section{Summary and Conclusions}
\label{conclusion}

We investigate the dependence of excess ultraviolet emission on rotation for evolved stars using {\it 2MASS} and {\it GALEX} broadband photometry in combination with APOGEE spectra. We define and measure $NUV$ excess for 133 unique giants 
in the $NUV-J$ versus $J-K_S$ color space. Our analysis 
shows that $NUV$ excess is strongly correlated with rotation with very high statistical significance.

We use our new empirical rotation-activity relation for red giants to explore the physical implications for the relationships between activity, stellar structure, and stellar evolution.
First, we find that the same rotation-activity relation applies for red giants in both the RGB and red clump stages of evolution, implying that rotation is by itself the dominant driver of activity over internal structure and/or sources of energy generation in evolved stars. 
Second, we find that our newly determined rotation-activity relation for cool giants is broadly similar to that observed in cool dwarfs, suggesting a common physical origin of activity across these very different stages of evolution. 
Interestingly, we find evidence that saturation of $NUV$ emission among rapidly rotating red giants in our sample also occurs at a similar level and at a similar Rossby number to that seen in M~dwarfs. 
Remarkably, we also find tentative evidence for so-called ``super-saturation" in our most rapidly rotating red giants, corresponding to stars rotating at a large fraction of breakup speed, as has also been suggested for extremely fast spinning M~dwarfs, although our sample size in this region is very small. 


Finally, we show, using the example of a recently reported red-giant/black-hole binary system, that our newly determined empirical rotation-activity relation for red giants can assist in characterizing red giants at present and forensically, in a broad array of astrophysical contexts. 
An interesting additional test for our relation that can be done in the future is to see if $NUV$-estimated $v\sin i$ distributions of lithium-rich giants are consistent with the rapid rotation fraction that has been predicted by \citet{2019arXiv190204102C}.

Most fundamentally, our findings suggest an underlying physical connection between rotation, {\bf convection} and activity for cool stars across the Hertzsprung-Russell diagram.

\acknowledgements
D.D.\ acknowledges partial funding support from NSF PAARE grant AST-1358862 through the Fisk-Vanderbilt Masters-to-PhD Bridge Program. J.T.\ acknowledges support provided by NASA through the NASA Hubble Fellowship grant No.~51424 awarded by the Space Telescope Science Institute, which is operated by the Association of Universities for Research in Astronomy, Inc., for NASA, under contract NAS5-26555.
K.G.S.\ acknowledges partial support from NASA grant 17-XRP17 2-0024.
{\bf We are grateful to the anonymous referee for a thorough and helpful review that substantially improved the paper.}

\bibliographystyle{aasjournal}
\bibliography{main} 

\end{document}